\documentclass[12pt]{article}

\begin{document}

\title{EPR-Bohr and Quantum Trajectories: \\
Entanglement and Nonlocality}

\author{Edward R. Floyd \\
Jamaica Village Road, Coronado, CA 92118-3208, USA \\
floyd@mailaps.org}

\date{\today}

\maketitle
\begin{abstract}

{Quantum trajectories are used to investigate the EPR-Bohr debate in a modern sense
by examining entanglement and nonlocality. We synthesize a single ``entanglement
molecule" from the two scattered particles of the EPR experiment. We explicitly
investigate the behavior of the entanglement molecule rather than the behaviors of
the two scattered particles to gain insight into the EPR-Bohr debate. We develop the
entanglement molecule's wave function in polar form and its reduced action, both of
which manifest entanglement. We next apply Jacobi's theorem to the reduced action to
generate the equation of quantum motion for the entanglement molecule to produce its
quantum trajectory.  The resultant quantum trajectory manifests entanglement and has
retrograde segments interspersed between segments of forward motion. This
alternating of forward and retrograde segments generates nonlocality and, within the
entanglement molecule, action at a distance. Dissection of the equation of quantum
motion for the entanglement molecule, while rendering the classical behavior of the
two scattered particles, also reveals an emergent ``entanglon" that maintains the
entanglement between the scattered particles. The characteristics of the entanglon
and its relationship to nonlocality are examined.}

\end{abstract}

\bigskip
 
\footnotesize


\bigskip


\clearpage

\normalsize
\section{Introduction}

``Can quantum mechanical description of physical reality be considered complete?"
was the title that Einstein, Podolsky and Rosen (EPR) [\ref{bib:epr}] and Bohr
[\ref{bib:bohr}] used in their 1935 debate regarding reality and completeness of
quantum mechanics. The issues circa 1935 were ``physical reality" and ``completeness
of the Schr\"odinger wave function, $\psi$." Subsequently, Bell [\ref{bib:bell}] and
the Aspect experiments [\ref{bib:aspect}] have shown quantum mechanics to be
nonlocal.   The modern issues of the EPR-Bohr debate are entanglement and
nonlocality [\ref{bib:fine},\ref{bib:hw}]. Many in the physics community remain
skeptical about the theoretical foundation for nonlocality in quantum mechanics not
withstanding the findings of experiments more accurate than the original Aspect
experiments with regard to detection and locality loopholes
[\ref{bib:hw}--\ref{bib:zeilinger}]. Herein, we investigate EPR phenomena with
quantum trajectories with a goal of answering the locality loophole issue. Quantum
trajectories are shown to render insight on nonlocality in quantum mechanics. In the
course of this investigation, analysis of the quantum trajectories revealed an
additional quantum entity, introduced as an ``entanglon", that can superluminally
maintain entanglement between the two EPR particles.

The quantum trajectory representation of quantum mechanics is a nonlocal,
phenomenological theory that is deterministic. Herein, ``deterministic" means in the
spirit of EPR that if without disturbing a system one can predict with certainty the
value of a physical quantity (the quantum trajectory), then there exists an element
of physical reality that corresponds to such a physical quantity (the quantum
trajectory) [\ref{bib:epr}].  Quantum trajectories with their nonlocal character are
adduced as a natural representation for investigating EPR phenomena and to render
insight into how entanglement induces nonlocality. The quantum Hamilton-Jacobi
equation underlies the quantum trajectory representation of quantum mechanics
[\ref{bib:prd34},\ref{bib:vigsym3}]. The underlying Hamilton-Jacobi formulation
couches the quantum trajectory representation of quantum mechanics in a
configuration space, time domain rather than a Hilbert space of wave mechanics.
Faraggi and Matone, using a quantum equivalence principle that connects all physical
systems by a coordinate transformation, have independently derived the quantum
stationary Hamilton-Jacobi equation (QSHJE) without using any axiomatic
interpretations of $\psi $ [\ref{bib:fm},\ref{bib:fm2}]. With Bertoldi, they have
extended their work to higher dimensions and to relativistic quantum mechanics
[\ref{bib:bfm}]. The quantum trajectory representation of quantum mechanics contains
more information then the Schr\"odinger wave function, $\psi$
[\ref{bib:prd34}--\ref{bib:fm2},\ref{bib:rc}--\ref{bib:fpl9}].  All of the foregoing
has posited the quantum trajectory representation as the superior method for
examining fundamental issues of quantum mechanics.  The quantum trajectory
representation has been used to investigate the foundations of quantum mechanics
free of axiomatic interpretations of Copenhagen philosophy
[\ref{bib:prd34}--\ref{bib:fp37a}]. With regard to the circa 1935 issue of
completeness of $\psi$, the quantum trajectory representation has already shown the
existence of microstates in $\psi$ that provides a counterexample showing that
$\psi$ is not an exhaustive description of quantum phenomena
[\ref{bib:prd34},\ref{bib:fm2},\ref{bib:rc}--\ref{bib:fpl9}].

This investigation studies the example considered by both EPR and Bohr in their 1935
papers where two identical particles without spin are entangled and scattered from
each other in opposite directions by some interaction
[\ref{bib:epr},\ref{bib:bohr}].  We investigate this example in a quantum
Hamilton-Jacobi representation and develop the quantum trajectory.  Rather than
examining the individual quantum trajectories of the two entangled particles, we
synthesize an ``EPR-molecule" from the two entangled particles and subsequently
examine the EPR-molecule's quantum trajectory to gain insight on how entanglement
induces nonlocality.  Synthesizing an EPR-molecule renders a reduced action in an
Euclidian space rather than in the configuration space described by the two
entangled particles. Synthesizing an EPR-molecule is reminiscent of synthesizing a
dispherical particle for an idealized quantum Young's diffraction experiment (a
simplified double slit experiment) where it was shown that the subsequent quantum
trajectory for the dispherical particle transited both slits simultaneously
[\ref{bib:fp37b}]. Quantum trajectory for multi-chromatic particles have also
explained wave packet spreading [\ref{bib:fp37a}].

The terminology ``EPR-molecule" is reserved for the example considered by both EPR
and Bohr in their 1935 papers where they examine identical particles recoiling in
opposite directions from each other after an entangling and scattering interaction
[\ref{bib:epr},\ref{bib:bohr}].  This investigation examines this situation in the
limit that the recoiling particles become identical.  For situations where the
recoiling particles are not identical, the terminology ``epr-molecule" is used. For
general situations of entanglement, the terminology ``entanglement molecule" is
used, which is a generalization of D\"{u}r's 2001 usage to describe entanglement
among qubits [\ref{bib:dur}].

Herein, the concept of a self-entangled, quantum particle
[\ref{bib:fp37b},\ref{bib:fp37a}] is extended to synthesize an entanglement molecule
from two entangled particles. Much of the formulation for describing EPR phenomenon
is common to that for self-entangled phenomenon, but the application differs
physically. Herein, we apply quantum trajectories to investigate the quantum motion
of entanglement molecules. For non-identical entangled particles, the  consequent
epr-molecule may spread and manifest nonlocality consistent with the Aspect
experiments [\ref{bib:aspect}]. Our investigation of EPR in a quantum trajectory
representation first synthesizes the epr-molecule. We next extract the generator of
the quantum motion (quantum reduced action or Hamilton's characteristic function)
for the epr-molecule from its wave function. Jacobi's theorem then renders the
quantum trajectory for the epr-molecule.  The resultant quantum trajectory has
retrograde segments interspersed between its segments of forward motion. This
alternating of forward and retrograde segments generates the nonlocality associated
with entanglement. Dissection of the equation of quantum motion for the quantum
trajectory reveals the classical motion for the two recoiling particles plus motion
for an emergent additional entity that contains the entanglement information. This
entity is designated as the ``entanglon". The entanglon is to the entangled molecule
what the chemical bond is to a standard molecule. The motion for the EPR-molecule is
determined from the motion for the epr-molecule in the limit that the recoiling
particles become identical.

In Section 2, we develop the formulation for applying quantum trajectories to the
EPR gedanken experiment. This includes synthesizing the entanglement molecule,
developing its generator of quantum motion, and subsequently developing its quantum
trajectory. In Section 3, we examine a particular example of an epr-molecule. We
generate its theoretical equation of quantum motion. In Section 4, we examine the
corresponding example for the EPR-molecule by taking the EPR-limit of our results
for the epr-example. In Section 5, we exhibit the emergence of the ``entanglon" in
the quantum trajectory representation of quantum mechanics. In Section 6, we present
conclusions.  Our conclusions include a discussion of how our quantum trajectory
representation differs with the Copenhagen interpretation and with Bohmian
mechanics.

\section{Formulation}

We adopt the physical setup of the original gedanken experiment considered by EPR
[\ref{bib:epr}] and Bohr [{\ref{bib:bohr}] for investigation.  However, we shall
examine the EPR experiment in a quantum Hamilton-Jacobi representation rather than
in a Schr\"{o}dinger wave function ($\psi$) representation to gain insight into the
relationship between entanglement and nonlocality. Let us consider two particles,
$P_1$ and $P_2$, with spatial wave functions , $\psi_1(x)$ and $\psi_2(x)$, that
interact through an instantaneous impulse (kick) at time $t=0$, rather than through
an interaction over the duration between $t=0,T$ as per EPR [{\ref{bib:epr}], and
then become entangled for $t>0$. The positions ($x_1,x_2$) of the two particles are
co-located at the time of impulse interaction, $t=0$ at $x_1,x_2=0$. The masses of
$P_1$ and $P_2$ are respectively given by $m$ and $\alpha^2m$ where $0 < \alpha \le
1$.  The factor $\alpha$ in $\psi_2$ is inserted arbitrarily as a convenient means
by which we may approach EPR in the limit $\alpha \to 1$.  More about this later.
For mathematical simplicity, let us conjure up some interaction and an inertial
reference system, reminiscent of EPR [\ref{bib:epr}] and Bohr[\ref{bib:bohr}], that
induces the two particles to recoil from each other in opposite directions after
impulse with their spatial wave functions given by

\begin{equation}
\psi_1(x)=\exp(ikx), \ \ \ \ \psi_2(x)=\alpha \exp(-ikx+i\beta); \ \ \ \ t > 0
\label{eq:recoil}
\end{equation}

\noindent where $k=[2(1+\alpha^2)mE]^{1/2}/\hbar$, $E$ is energy of the epr-molecule
and $-\pi<\beta \le \pi$. The term $\beta$ represents a phase shift between the two
particles.  In our chosen reference system, $(\psi_1,\psi_2)$ form a set of
independent solutions of the Schr\"{o}dinger equation with energy $E$
[\ref{bib:prd34},\ref{bib:fm2}], which helps facilitate the application of the
quantum trajectory representation. The wave functions $\psi_1$ and $\psi_2$ have not
been normalized absolutely but do have relative normalization with regard to each
other as manifested by the factor $\alpha$. By Eq.\ (\ref{eq:recoil}), $\psi_1$ and
$\psi_2$ are not the wave functions for identical particles unless $\alpha = 1$. EPR
[\ref{bib:epr}] and Bohr [\ref{bib:bohr}] considered identical particles.  For
completeness, the particular combination of impulse interaction at $t=0$ and
particle velocities $\dot{x}_1$ and $\dot{x}_2$ for $-1 \ll t<0$ necessary to render
the particular results of Eq. (\ref{eq:recoil}) is generally not unique.

While the particles $P_1$ and $P_2$ have causal positions $x_1$ and $x_2$
respectively, their wave functions $\psi_1$ and $\psi_2$ in the Copenhagen
interpretation render the Born probability amplitude over $x$. In the quantum
trajectory representation, the set $(\psi_1,\psi_2)$ of independent solutions to the
Schr\"{o}dinger equation in one dimension are related to the reduced action in the
quantum trajectory representation of quantum mechanics through the invariance of the
Schwarzian derivative of the reduced action with regard to $x$ under a M\"{o}bius
transform $(a\psi_1-b\psi_2)/(c\psi_1-d\psi_2),\ ad-bc \ne 0$ [\ref{bib:fm2}].

Our criterion for choice of inertial reference system, for which  $\psi_1$ and
$\psi_2$ have the wave numbers $k$ and $-k$, generates the relationship
$x_1=-x_2/\alpha^2$ for the positions of the two particles. This is an extension of
Fine's conservation of relative position [\ref{bib:fine}]. For $\alpha=\pm 1$,
conservation of relative position holds, and the inertial reference system is the
center-of-mass inertial system. Conservation of relative position will induce loss of 
parameter independence and outcome independence [\ref{bib:hw}] in the EPR
experiment.

EPR and Bohr assumed that, for times sufficiently long after interaction at $t=0$,
then $x_1+x_2 \gg 1$ sufficiently to ensure ``separability" of the particles $P_1$
and $P_2$. But we herein assume that the two particles remain entangled no matter
how far apart they become as first confirmed by the Aspect experiments
[\ref{bib:aspect}].

For entanglement in the wave function representation of quantum mechanics, we may
synthesize an epr-molecule as a simple polar wave function, $\psi_{epr}$, from the
entangled pair (bipolar wave function) , $\psi_1$\ and $\psi_2$, by
[\ref{bib:fp37a},\ref{bib:holland}]

\begin{eqnarray}
\psi_{epr}(x) & = & \overbrace{\exp(ikx) + \alpha \exp(-ikx +
i\beta )}^{\mbox {\normalsize bipolar wave function}}  \nonumber \\
        & = & \underbrace{[1+\alpha^2+2 \alpha \cos(2kx+\beta)]^{1/2} \exp \left[ i \arctan
        \left( \frac{\sin(kx) - \alpha \sin(kx+\beta)}{\cos(kx) + \alpha
\cos(kx+\beta)}\right) \right]}_{\mbox {\normalsize polar wave function is still an
eigenfunction for $E=\hbar^2k^2/[2m(1+\alpha^2)]$.}} \label{eq:eprpsi}
\end{eqnarray}

\noindent where we have dropped the subscript upon particle position $x$ by the
extension of conservation of relative position. The above construction is just
superpositional principle at work. It converts a bipolar {\itshape Ansatz} to a
polar {\itshape Ansatz}
[\ref{bib:vigsym3},\ref{bib:fm2},\ref{bib:prd26},\ref{bib:prd25}--\ref{bib:wyatt}].
In the EPR limit, then $\lim_{\alpha \to 1} \psi_{epr} \to \psi_{EPR} = 2
\cos(kx),i2 \sin(kx)$ respectively for $\beta = 0,\pi$ as expected. From Eq.\
(\ref{eq:eprpsi}), $\psi_{epr}$ has the same form as a dichromatic wave function
$\psi_{dichromatic}$ investigated in Reference \ref{bib:fp37a}. But $\psi_{epr}$ and
$\psi_{dichromatic}$ represent different physics as the two spectral components of
$\psi_{dichromatic}$ induce self-entanglement within a dichromatic particle.

The wave function for the epr-molecule, $\psi_{epr}$, as exhibited by Eq.\
(\ref{eq:eprpsi}), does not uniquely specify the epr-components.  For example, the
entanglement of a running wave function, $(1-\alpha)\exp(ikx)$ and a standing wave
function, $2 \alpha \exp(-i\beta/2) \cos(kx+\beta/2)$ would also render the very
same $\psi_{epr}$ given by Eq.\ (\ref{eq:eprpsi}).  By the superpositional theorem,
$\psi_{epr}$ remains valid for any combination of particles as long as the
collective sum of their spectral components are consistent with the right side of
the upper line of Eq.\ (\ref{eq:eprpsi}).

In the wave function representation, $\psi_{epr}$ as represented by Eq.\
(\ref{eq:eprpsi}), is inherently nonlocal for it is not factorable, that is
[\ref{bib:ch}] $\psi_{epr} \ne K \psi_1 \psi_2$ where K is a constant. Any
measurement upon the $\psi_{epr}$ for the epr-molecule concurrently measures
$\psi_1$ and $\psi_2$. Likewise, in the quantum trajectory representation of quantum
mechanics, entanglement implies that the reduced action (Hamilton's characteristic
function) for the epr-molecule, $W_{epr}$ is inseparable by particles, that is
$W_{epr} \ne W_{\mbox{\scriptsize particle 1}} + W_{\mbox{\scriptsize particle 2}}$.

The $\psi_{epr}$ is not the wave function representing EPR landscape. The actual
wave function for the EPR-molecule, $\psi_{EPR}$, for identical particles is given
by

\[
\psi_{EPR} = \lim_{\alpha \to 1} (\psi_{epr}).
\]

\noindent In general, we shall investigate EPR phenomena, where $\alpha=1$, by

\[
\lim_{\alpha \to 1} {\Big(\mbox{epr-phenomenon}\Big) \to \mbox{EPR-phenomenon}}.
\]

\noindent This avoids directly working with standing waves to establish quantum
trajectories and permits us to study the behavior of quantum trajectories and other
phenomena in the limit that the complex running wave function, $\psi_{epr}$, becomes
a real standing wave function, $\psi_{EPR}$, as $\alpha \to 1$.

A generator of the motion for the epr-molecule is its reduced action, $W_{epr}$. Its
reduced action may be extracted from the un-normalized $\psi_{epr}$ as microstates
do not exist for $\psi_{epr}$ [\ref{bib:fpl9}].  The reduced action is given by
[\ref{bib:fp37a},\ref{bib:holland}]

\begin{equation}
W_{epr} = \hbar \arctan \left( \frac{\sin(kx) - \alpha \sin(kx+\beta)}{\cos(kx) +
\alpha \cos(kx+\beta)}\right). \label{eq:eprw}
\end{equation}

\noindent Whereas we extracted the reduced action, $W_{epr}$, from the Schr\"odinger
wave function herein for convenience, Faraggi and Matone have shown  that in general
the reduced action may be derived from their quantum equivalence  principle
independent of the Schr\"odinger formulation of quantum mechanics [\ref{bib:fm}].
The reduced action, $W_{epr}$, is also the solution of the QSHJE for
$E=\hbar^2k^2/[2m(1+\alpha^2)]$ [\ref{bib:fp37a}].  Equation (\ref{eq:eprw}) posits
a deterministic $W_{epr}$ in Euclidean space in contrast to $\psi_{epr}$ with its
probability amplitude being posited in Hilbert space. The absolute value of
$W_{epr}$ increases monotonically with $x$ as the arctangent function in $W_{epr}$
as it jumps to the next Riemann sheet whenever the the underlying tangent function
becomes singular.

The conjugate momentum for epr-molecule is given by

\begin{equation}
\partial W_{epr}/\partial x = \frac{\hbar k}{[1+\alpha^2+2
\alpha \cos(2kx+\beta)]}. \label{eq:cm}
\end{equation}

\noindent The conjugate momentum manifests entanglement by the cosine term in the
denominator on the right side of Eq.\ (\ref{eq:cm}).  We note from Eqs.\
(\ref{eq:cm}) and (\ref{eq:eom}) that the conjugate momentum is not the mechanical
momentum, i.e.,

\begin{equation} \partial W_{epr}/\partial x \ne m\dot{x}.
\label{eq:mm} \end{equation}

The equation of quantum motion for the epr-molecule is generated from $W_{epr}$ by
Jacobi's theorem as

\begin{equation}
\underbrace{t_{epr}-\tau=\frac{\partial W_{epr}}{\partial E}}_{\mbox{Jacobi's
theorem}}=\frac{mx(1-\alpha^2)}{\hbar k[1+\alpha ^2 +2\alpha \cos(2kx + \beta)]}
\label{eq:eom}
\end{equation}

\noindent where $t$ is time and $\tau$ specifies the epoch. The quantum trajectory
for the epr-molecule is in Euclidean space and renders determinism as proposed by
EPR [\ref{bib:epr}] for the position of the epr-molecule as a function of time that
can be predicted with certainty without disturbing the system. In the forgoing, we
note that our use of ``certainty" is appropriate for three reasons. First, in the
Copenhagen interpretation, the Heisenberg uncertainty principle uses an insufficient
subset of initial values of the necessary and sufficient set of initial values that
specify unique quantum motion [\ref{bib:fm2},\ref{bib:prd29},\ref{bib:ijmpa15}].
Second, the quantum trajectories exist in Euclidean space here while the
Schr\"{o}dinger wave function representation is formulated in Hilbert space
[\ref{bib:rc}].  And third, the quantum Hamilton-Jacobi representation contains more
information than  the Schr\"odinger wave function representation and renders a
unique, deterministic quantum trajectory
[\ref{bib:fm2},\ref{bib:prd29},\ref{bib:fpl9},\ref{bib:ijmpa15}]. Realism follows
from determinism for the epr-molecule maintains a precise, theoretical quantum
trajectory independent of it being measured.  Nevertheless, nothing herein implies
that a measurement on an epr-molecule does not physically disturb the epr-molecule
in compliance with Bohr's complementarity principle.

The use of Jacobi's theorem to develop an equation of quantum motion, Eq.\
(\ref{eq:eom}), is consistent with Peres's quantum clocks where $t-\tau=\hbar
(\partial \varphi /\partial E)$ where $\varphi$ is the phase of the complex wave
function of the particle under consideration [\ref{bib:peres}]. Equation
(\ref{eq:eom}) is a generalization of this for it applies to situations where the
wave function is real [\ref{bib:fm2},\ref{bib:prd26},\ref{bib:fpl9}].

We also note that the development of quantum trajectories differs from those of
Bohmian mechanics [\ref{bib:bohm}].  Bohmian mechanics assumes that the conjugate
momentum, $\partial W_{epr}/\partial x$, is the mechanical momentum in contradiction
to Eq.\ (\ref{eq:mm}) and subsequently integrates it to render an equation of
quantum motion that differs from Eq.\ (\ref{eq:eom}).  Recently, Ghose has shown for
some entangled multiparticle systems that choosing the particle distribution in
Bohmian mechanics consistent with the ``quantum equilibrium hypothesis" cannot be
assured: a Bohmian interpretation becomes problematic for such systems
[\ref{bib:ghose}].  Ghose did investigate in a Bohmian representation the
entanglement, Eq.\ (\ref{eq:eprpsi}), which is studied herein.

In closing this section, we note that measurements on $\psi_{epr}$ concurrently
measure $\psi_1$ and $\psi_2$ support the position of Bohr in the EPR-Bohr debates
[\ref{bib:bohr}]. On the other hand, the very existence of quantum trajectories for
the epr-molecule supports the position of EPR with regard to reality
[\ref{bib:epr}]. As previously discussed, the quantum Hamilton-Jacobi representation
contains more information than $\psi$ which challenges the completeness of $\psi$
which also supports the position of EPR.

\section{Example}

Let us consider the particular example of the quantum trajectories of an
epr-molecule specified by $\hbar=1, \ m=1, \ k=\pi/2, \ \alpha = 0.5$, \ $\tau=0$,
and $\beta = 0,\pi$. The resultant quantum trajectories, which are governed by Eq.\
(\ref{eq:eom}), are exhibited on Fig.\ 1 where the solid line renders the quantum
trajectory for $\beta=0$ and dashed line, $\beta=\pi$. These quantum trajectories
are launched from the origin, $(t,x)=(0,0)$. Near $x = 1,2,3,\cdots$, the quantum
trajectory for the epr-molecule with $\beta = 0$ on Fig.\ 1 has turning points with
regard to time, $t$, where the quantum trajectories change between forward and
retrograde motion [\ref{bib:fp37a}]. The turning points cause the quantum trajectory
to alternate between forward and retrograde motion implying nonlocality and action
at a distance as the quantum trajectory at various instances of time has separate,
multiple locations. Furthermore, the good behavior (at least continuous first-order
derivatives) implies superluminality of the quantum trajectories at the turning
points for the epr-molecule where $\dot{x} \to \pm \infty$ at the extrema in $t$.
This superluminality is another manifestation of nonlocality.  We note that these
superluminalities at the turning points are integratable as exhibited on Fig.\ 1.

The quantum trajectory for the epr-molecule, as exhibited by Fig.\ 1, is restricted
to the approximate wedge given by

\[
\frac{mx}{3\hbar k} \le t \le \frac{3mx}{\hbar k}.
\]

\noindent The upper boundary of the wedge, $t_u=3mx/(\hbar k)$, manifests maximum
destructive interference between $\psi_1$ and $\psi_2$ while the lower boundary,
$t_{\ell}=mx/(3 \hbar k)$, manifests maximum reinforcement. This may be generalized
with regard to $\alpha$ by

\begin{equation}
\frac{(1-\alpha)mx }{(1+\alpha)\hbar k} \le t \le
\frac{(1+\alpha)mx}{(1-\alpha)\hbar k}. \label{eq:wedge}
\end{equation}

\noindent  This wedge may be densely filled by varying the phase shift $\beta$ over
its range $(-\pi/2,\pi/2)$ as exhibited by Fig.\ 2. For $\alpha \ll 1$, latent early
time reversals may be suppressed [\ref{bib:fp37a}].

As the quantum trajectory for the epr-molecule progresses out the wedge away from
its launch point at the wedge's apex at the origin $(t,x)=(0,0)$, the durations of
time spent on individual forward and retrograde segments increase.  The dichromatic
particle offers a precedent for understanding this motion of alternating forward and
retrograde segments whose duration progressively increase as manifesting wave packet
spreading [\ref{bib:fp37a}]. Here, the analogous behavior for the epr-molecule
manifests an increasing spatial displacement between its two component particles,
$P_1$ and $P_2$.

There is another way to interpret the quantum trajectories exhibited in Figs.\ 1 and
2 where the concept of retrograde motion is replaced by invoking the use of creation
and annihilation operations at the turning points [\ref{bib:fp37a}]. At the local
temporal minima, there is maximum reinforcement between $\psi_1$ and $\psi_2$, which
synthesize $\psi_{epr}$, at these temporal local minima where pairs of quantum
trajectories for the epr-molecule are spontaneously created. Within each pair, one
quantum trajectory propagates in the $+x$ direction; the other, in the $-x$
direction. Note that these creation operations do not imply that $\psi_{epr}$ has
been spectrally analyzed into $\psi_1$ and $\psi_2$ to propagate separately on the
two different branches: rather the creation operations manifest spontaneous
nonlocality where $\psi_{epr}$ propagates along both branches. Each branch of the
pair terminates at local temporal maximum where it is annihilated along with another
branch from another pair of quantum trajectories as exhibited on Fig.\ 1. These
annihilated quantum trajectories were created at different local temporal minima and
propagate in opposite directions with regard to $x$. The local temporal maxima
represent points of maximum interference between $\psi_1$ and $\psi_2$ within
$\psi_{epr}$.

\section{Quantum trajectory for EPR-molecule}

The wave function for the EPR-molecule is a standing wave function.  As such, its
corresponding quantum trajectory is ill defined.  We shall resolve its quantum
trajectory by a limiting process. We still assume the conditions $\hbar =1,\ m=1, \
k=\pi/2,$ and $\beta=0$.  For $\beta=0$, the epr-reduced action simplifies to

\[
 W_{epr}=\hbar \left[ \arctan \left(\frac{1-\alpha}{1+\alpha} \tan(kx) \right)
 \right].
 \]

 The EPR wave function by the upper line of Eq.\ (\ref{eq:eprpsi}) with $\alpha = 1$
trivially represents a standing wave function, $2 \cos(kx)$.  Likewise, the limiting
process, $\alpha \to (1-)$, when applied to the second line of Eq.\
(\ref{eq:eprpsi}), also renders

\begin{equation}
\lim _{\alpha \to (1-)} \big(\psi _{epr}\big) = 2\cos(kx) = \psi_{EPR}.
\label{eq:psiEPR}
\end{equation}

\noindent Our limiting process for EPR has $\alpha$ approach 1 from below, $\alpha
\to (1-)$. Concurrently, the instantaneous inertial reference frame, which is
dependent upon $\alpha$, is continuously constrained throughout the limiting process
to maintain the wave numbers, $k$ and $-k$ for $\psi_1$ and $\psi_2$ respectively.
In the limit $\alpha \to (1-)$, both edges of the wedge exhibited in Figs.\ 1 and 2
become orthogonal [\ref{bib:fp37a}]. The wedge spans the entire quadrant $t,x \ge 0$
of the $t,x$-plane.  Had we chosen to take the limit of $\alpha$ approaching 1 from
above, then Eq.\ (\ref{eq:psiEPR}) would still be valid but the wedge would have
spanned the quadrant $t \ge 0,x \le 0$ of the $t,x$-plane.

The equation of quantum motion for the EPR-molecule, which by Jacobi's theorem,
$t_{EPR}=\partial W_{EPR}/\partial E$, is rendered by taking the limit of $\alpha
\to 1$ from below of the epr equation of quantum motion, Eq.\ (\ref{eq:eom}). For a
launch point (initial position) of $(t,x)=(0,0)$, quantum motion for the
EPR-molecule in the limit $\alpha \to (1-)$ is given by [\ref{bib:fpl9}]

\begin{equation}
\lim _{\alpha \to (1-)} t_{epr}= t_{EPR} = \sum _{n=1}^{\infty}\delta [x-(2n-1)\pi
/(2k)] = \sum _{n=1}^{\infty}\{ \delta [x-(2n-1)] , \ \ x>0, \ \tau_{EPR}=0
\label{eq:eom-}
\end{equation}

\noindent  consistent with the equation of quantum motion, Eq.\ (\ref{eq:eom}). For
$x<0$ and the launch point still at $(t,x)=(0,0)$, we investigate the case $1 \le
\alpha \le \infty$ using the limiting process $\alpha \to 1$ from above. This
renders

\begin{equation}
\lim _{\alpha \to (1+)} t_{\downarrow epr} = t_{\downarrow EPR} = -\sum
_{n=1}^{\infty} \delta [x-(2n-1)\pi /(2k)] = -\sum _{n=1}^{\infty}\{ \delta
[x-(2n-1)], \ \ x<0, \ \tau_{\downarrow EPR}=0 \label{eq:eom+}
\end{equation}

\noindent where the prefix $\downarrow$ in the subscripts denotes the limiting
process $\alpha \to (1+)$ to generate quantum trajectories into the domain $x<0$.
The prefix $\uparrow\hspace{-4pt}\cup\hspace{-5pt}\downarrow$ denotes the union of
the limiting processes $\alpha \to (1\mp)$.  For launch point at $x=0$,
$\uparrow\hspace{-4pt}\cup\hspace{-5pt}\downarrow$EPR-molecule has positive infinite
velocity for $x>0$ and $x \ne 1,3,5,\cdots$ by Eq.\ (\ref{eq:eom-}) and negative
infinite velocity for $x<0$ and $x \ne -1,-3,-5,\cdots$ by Eq.\ (\ref{eq:eom+}) in
this nonrelativistic representation. These infinite magnitudes of velocity at  $x
\ne \pm 1,\pm 3,\pm 5,\cdots$ imply action at \emph{infinite} distances within the
EPR-molecule in this nonrelativistic examination. Also, at the trigger points of the
$\delta$-function of Eqs.\ (\ref{eq:eom-}) and (\ref{eq:eom+}), $x= \pm 1,\pm 3,\pm
5,\cdots$, the EPR-molecule has nil velocity consistent with $\psi_{EPR}=2
\cos(kx)$. Thus, the limiting process, $\alpha \to 1$, renders the expected standing
wave function for $\psi_{EPR}$ given by Eq.\ (\ref{eq:psiEPR}) while the limiting
process also renders a consistent equation of quantum motion for
$t_{\uparrow\hspace{-1pt}\cup\hspace{-1pt}\downarrow}$ given by Eqs.\
(\ref{eq:eom-}) and (\ref{eq:eom+}).

The alternative interpretation using creation and annihilation operations, which
already has been discussed in Section 3, begs the question whether these operations
imply high energy processes.  They do not. This is shown by applying Faraggi and
Matone's effective quantum mass, $m_{Q_{EPR}}=m(1-\partial Q_{EPR}/\partial E)$
where $Q$ is Bohm's quantum potential [\ref{bib:fm}], to this investigation. For the
EPR-molecule, $m_{Q_{EPR}}$ becomes [\ref{bib:fp37a}]

\begin{eqnarray}
\lim_{\alpha \to 1} m_{Q_{\pm EPR}} & = & 0, \ \ x \ne \pm 1,\pm 3,\pm 5, \cdots \nonumber \\
                        & = & \infty, \ \ x = \pm 1,\pm 3,\pm 5, \cdots.
                              \label{eq:EPReom}
\end{eqnarray}

\noindent  Note that $m_{Q_{\pm EPR}}$  here becomes infinite where the velocity of
the EPR-molecule is nil and becomes nil where the velocity is infinite. This is
consistent with conjugate momentum remaining finite [\ref{bib:fm}].  Herein, neither
do creation operations imply endoergic processes nor do annihilation operations
imply exoergic processes.

\section{The ``entanglon"}

\noindent Let us now demonstrate the emergence of the entanglon for an epr-molecule
from the equation of quantum motion in the quantum trajectory representation of
quantum mechanics. We shall dissect the equation of quantum motion for the synthetic
epr-molecule, Eq. (\ref{eq:eom}), to resolve the contributions to $t_{epr}$ by
particles $P_1$ and $P_2$ individually. These two individual contributions are
insufficient by themselves to render $t_{epr}$ for there remains a contribution due
to the entanglement between the two particles. Equation (\ref{eq:eom}) may be
dissected as

\begin{eqnarray}
t_{epr} & = & \ \frac{mx(1-\alpha^2)}{\hbar k[1+\alpha ^2 +2\alpha \cos(2kx +
\beta)]} \nonumber \\
& = & \ \underbrace{\frac{mx}{\hbar k} \frac{1}{1+\alpha^2}}_{\mbox{\small particle
1}} \ - \ \underbrace{\frac{mx}{\hbar k} \frac{2\alpha \frac{1-\alpha^2}{1+\alpha^2}
\cos(2kx+\beta)}{1+\alpha^2+2\alpha \cos(2kx+\beta)}}_{\mbox{\small entanglon}} \ -
\ \underbrace{\frac{mx}{\hbar k} \frac{\alpha^2}{1+\alpha^2}}_{\mbox{\small particle
2}} \label{eq:aeom}
\end{eqnarray}

\noindent where the epoch has been set as $\tau=0$.  The contributions to $t_{epr}$
from particles 1 and 2 are weighted. In the EPR limit, $\alpha \to 1$, the
contributions of particles 1 and 2 cancel each other. The remaining contribution
that emerges
 in Eq.\ (\ref{eq:aeom}) has been allocated to an entity now identified as the
``entanglon".  Its contribution to $t_{epr}$ in Eq.\ (\ref{eq:aeom}) is identified
as $t_{epr_e}$ Then, in the EPR limit, $\alpha \to (1-)$, $t_{EPR_e}$ is given by

\begin{eqnarray}
t_{EPR_e} & = & \lim_{\alpha \to (1-)} \left( \frac{mx}{\hbar k} \frac{2\alpha
\frac{1-\alpha^2}{1+\alpha^2} \cos(2kx+\beta)}{1+\alpha^2+2\alpha \cos(2kx+\beta)}
\right) \nonumber \\
              & = & \left\{ \begin{array}{ll}
                        0,\ \ \ x \ne 1,3,5,\cdots \\
                         \lim_{\alpha \to (1-)}\left( \frac{mx}{\hbar k}
\frac{2}{1-\alpha}\right) \to \infty,\ \ \ x=1,3,5,\cdots.
\end{array}                    \right.
\label{eq:teEPR}
\end{eqnarray}

\noindent  Hence, $t_{EPR}$ exhibits multi $\delta$-function behavior at $x = n\pi
/(2k), n=1,3,5,\cdots .$  As the contributions to $t_{EPR}$ from particles 1 and 2
mutually cancel each other in the EPR limit $\alpha \to (1-)$ as shown by Eq.\
(\ref{eq:aeom}), we have that $t_{EPR}=t_{EPR_e}$. The $\delta$-function behavior of
$t_{EPR_e}$ for the entanglon as given by Eq.\ (\ref{eq:teEPR}) is consistent with
the motion of the standing wave exhibited by Eq.\ (\ref{eq:EPReom}) at
$x=1,3,5,\cdots$ for $\beta=0$. Thus, the entanglon induces retrograde motion, which
manifests nonlocality.  The entanglon in the EPR limit implies action (entanglement) at infinite
distances within the EPR-molecule as $t_{EPR_e} \to 0$ by Eq.\ (\ref{eq:teEPR}) for
$x=1,3,5,\cdots$.

The entanglon is not an ``external" force carrier between particles such as the photon, graviton, etc.\ for the latent
motions for the individual particles $P_1$ and $P_2$ of the epr-molecule remain linear with constant velocity 
as shown in Eq.\ (\ref{eq:eom}). Nor does the entanglon change either wave function,
$\psi_1$ or $\psi_2$. Nevertheless, the entanglon does maintain the correlation between
$\psi_1$ and $\psi_2$, which it may do so superluminally.  In so doing, the entanglon renders an ``internal" force within the epr-molecule
influencing the quantum trajectory of the epr-molecule while maintaining a coherent epr-molecule.

The entanglon also has characteristics in common with the gluon. Neither exists in
isolation. When coherence within the epr-molecule is lost, then the entanglon no
longer exists. There is another characteristic in common for entanglons and gluons
which regards strength with range.  Gluons become stronger with range. Also, as
range increases, the entanglon, as well as the epr-molecule, spontaneously develops
an additional pair of segments that alternate with regard to retrograde and forward
motion in the quantum trajectory. These segments imply the existence of multipaths,
which are inherently nonlocal, for the entanglon. The number of multipaths increase
with range, which mitigates any loss of coherence between the entangled particles
with range. Thus, the concept that entangled particles that are widely separated in
this nonrelativistic investigation should become independent of each other due to
Einstein locality is refuted.

For completeness, the forward and retrograde segments of the entanglon are
reminiscent of Cramer's transactional interpretation of quantum mechanics
[\ref{bib:cramer}]. The transactional interpretation postulates that a quantum
interaction be a standing wave synthesized from a retarded (forward-in-time) wave
and an advanced (retrograde) wave.

The concept of the entanglon also supports a hierarchy of entanglement critical for
an undivided universe that has been postulated in Bohmian mechanics [\ref{bib:bh}].

\section{Conclusions}

We conclude that entangled particles may be synthesized into entanglement molecules.
The quantum trajectory representation of quantum mechanics does describe causal
behavior of the entanglement molecule without invoking the Born probability
postulate for $\psi$. The particular quantum trajectory for an entanglement molecule
may be specified by a single constant of the motion, $E$. The quantum trajectory
representation including Faraggi and Matone's quantum equivalence principle and
their quantum effective mass does resolve some of the mysteries of EPR.  The quantum
trajectory representation renders the emergence of the entanglon which maintains
coherence between widely separated, entangled entities.

Quantum trajectories in a nonrelativistic theory have shown for the EPR gedanken
experiment that entanglement may be maintained superluminally. In the case of the
EPR limit, entanglement is maintained instantaneously.  Also, quantum trajectories
in the EPR limit imply action at infinite distances in this nonrelativistic
investigation. Hence, the locality loophole cannot be closed.

This opus is consistent with Copenhagen through the description of the wave function
for the epr-molecule, $\psi_{epr}$ as exhibited by Eq.\ (\ref{eq:eprpsi}) but
differs thereafter. The anticipated Copenhagen response would stipulate that a
measurement upon $\psi_{epr}$ would render a probabilistic outcome for the
epr-molecule.  By axiomatic precept, Copenhagen has denied the very existence of the
deterministic quantum trajectories, which were used herein. As noted in the
Introduction, the quantum trajectory interpretation of quantum mechanics has already
shown that $\psi$ is not a complete description of quantum phenomena
[\ref{bib:fm2},\ref{bib:rc}--\ref{bib:fpl9}].

This opus is also consistent with Bohmian mechanics [\ref{bib:holland}] through the
description of the epr-reduced action, $W_{epr}$ as exhibited by Eq.\
(\ref{eq:eprw}) but differs thereater. Both representations are based upon the same
quantum Hamilton-Jacobi equation and develop the same generator of quantum motion.
While $W_{epr}$ is a common generator of quantum motion, quantum trajectories and
Bohmian mechanics have however different equations of quantum motion.  The quantum
trajectory representation develops its equation of quantum motion from Jacobi's
theorem, Eq.\ (\ref{eq:eom}). On the other hand, the Bohmian equation of quantum
motion is the integration of the conjugate momentum, Eq.\ (\ref{eq:cm})
[\ref{bib:holland}].

For completeness, should a measuring process on the entangled molecule use a matched
filter designed to measure some property of $\psi_1$ for example, then the measuring
process will detect that property of $\psi_1$.  To detect the entangled molecule,
the measuring filter must be matched to the entanglement molecule as a whole.

\bigskip

\noindent {\bf Acknowledgements}

\bigskip

I heartily thank Marco Matone and Alon E.\ Faraggi for their interesting discussions
and encouragement.

\bigskip

\clearpage

\noindent {\bf References}

\begin{enumerate}\itemsep -.06in

\item \label{bib:epr} A.\ Einstein, B.\ Podolski and N.\ Rosen, Phys.\ Rev.\
{\bfseries 47}, 777 (1935).

\item \label{bib:bohr} N.\ Bohr, Phys.\ Rev.\ {\bfseries 48}, 696 (1935).

\item \label{bib:bell} J.\ Bell, {\it Speakable and Unspeakable in Quantum
Mechanics} (Cambridge University Press, Cambridge, 1987).

\item \label{bib:aspect} A.\ Aspect, P.\ Grangier and G.\ Roger, Phys. Rev.\ Lett.\
{\bfseries 47}, 460 (1981); {\bfseries 49}, 91 (1982); A.\ Aspect, J.\ Dalibard and
G.\ Roger, Phys.\ Rev.\ Lett.\ {\bfseries 49}, 1804 (1982).

\item \label{bib:fine} A.\ Fine,
$<$http://plato.stanford.edu/archives/sum2004/entries/qt-epr/$>$.

\item \label{bib:hw} D.\ Home and A.\ Whitaker,  {\it Einstein's Struggles with
Quantum Theory} (Springer, New York, 2007).

\item \label{bib:kwiat} P.\ G.\ Kwiat, K.\ Mattle, H.\ Weinfurter, A.\ Zeilinger,
A.\ V.\ Sergienko, and Y.\ H.\ Shih, Phys.\ Rev.\ Lett.\ {\bfseries 75}, 4337
(1995).

\item \label{togerson} J.\ R.\ Togerson, D.\ Branning, C.\ H.\ Monken, and L.\
Mandel, Phys.\ Lett.\ {\bfseries A 204}, 323 (1995).

\item \label{digiuesppe} G.\ Di Giuseppe, F.\ De Martini, and D.\ Boschi, Phys.\
Rev.\ {\bfseries A 56}, 176 (1997).

\item \label{hardy} D.\ Boschi, S.\ Branca, F.\ De Martini, and L.\ Hardy, Phys.\
Rev.\ Lett.\ {\bfseries 79}, 2755 (1997).

\item \label{bib:weihs} G.\ Weihs, T.\ Jennewein, C.\ Simon, H.\ Weinfurter, and A.\
Zeilinger, Phys.\ Rev.\ Lett.\ {\bfseries 81}, 5039 (1998).

\item \label{bouwmeester} D.\ Bouwmeester, J.-W.\ Pan, M.\ Daniell, H.\ Weinfurter,
and A.\ Zeilinger, Phys.\ Rev.\ Lett.\ {\bfseries 82}, 1345 (1999).

\item \label{bib:tittel} W.\ Tittel, J.\ Brendel, N.\ Gisin, and H.\ Zbinden, Phys.\
Rev.\ {\bfseries A 59}, 4150 (1999).

\item \label{bib:rowe} M.\ A.\ Rowe, D.\ Klepinski, V.\ Meyer, C.\ A.\ Sackett, V.\
M.\ Itano, C.\ Monroe, and D.\ J.\ Wineland, Nature {\bfseries 409}, 791 (2001).

\item \label{bib:zeilinger} S.\ Gr\"oblacher, T.\ Paterek, R.\ Kaltenbaek, C.\
Brukner, M.\ \'Zukowski, M.\ Aspelmeyer, and A.\ Zeilinger, Nature {\bfseries 446},
1469 (2007).

\item \label{bib:prd34} E.\ R.\ Floyd, Phys.\ Rev.\ {\bf D 34}, 3246 (1982).

\item \label{bib:vigsym3} E.\ R.\ Floyd, {\it Gravitation and Cosomology: From the
Hubble Radius to the Planck Scale; Proceedings of Symposium in Honour of the 80th
Birthday of Jean-pierre Vigier}, ed. by R. L. Amoroso et al (Kluwer Academic, 721
Dordrecht, 2002), extended version quant-ph/00009070.

\item \label{bib:fm} A.\ E.\ Faraggi and M.\ Matone, Phys.\ Rev.\ Lett.\ {\bf 78},
163 (1997) hep-th/9606063; Phys.\ Lett.\ {\bf B 437}, 369 (1997), hep-th/9711028;
{\bf B 445}, 77 (1999), hep-th/9809125; 357 (1999), hep-th/9809126; {\bf B 450}, 34
(1999),  hep-th/9705108; {\bf A 249}, 180 (1998), hep-th/9801033.

\item \label{bib:fm2} A.\ E.\ Faraggi and M.\ Matone, Int.\ J.\ Mod.\ Phys.\ {\bf A
15}, 1869 (2000) hep-th/98090127.

\item \label{bib:bfm} G.\ Bertoldi, A.\ E.\ Faraggi and M.\ Matone, Class.\ Quant.\
Grav.\ {\bfseries 17} 3965 (2000), hep-th/9909201.

\item \label{bib:rc} R.\ Carroll, Can.\ J.\ Phys.\ {\bf 77}, 319 (1999),
quant-ph/9904081; {\it Quantum Theory, Deformation and Integrability} (Elsevier,
2000, Amsterdam) pp. 50--56, {\it Uncertainty, Trajectories, and Duality},
quant-ph/0309023.

\item \label{bib:prd26} E.\ R.\ Floyd,  Phys.\ Rev.\ {\bf D 26}, 1339 (1982).

\item \label{bib:prd29} E.\ R.\ Floyd,  Phys.\ Rev.\ {\bf D 29}, 1842 (1982).

\item \label{bib:fpl9} E.\ R.\ Floyd, Found.\ Phys.\ Lett.\ {\bf 9}, 489 (1996),
quant-ph/9707051.

\item \label{bib:fp37b} E.\ R.\ Floyd, Found.\ Phys.\ {\bfseries 37}, 1403 (2007),
quant-ph/0605121.

\item \label{bib:fp37a} E.\ R.\ Floyd, Found.\ Phys.\ {\bfseries 37}, 1386 (2007),
quant-ph/0605120.

\item \label{bib:dur} W.\ D\"{u}r, Phys.\ Rev.\ {\bfseries A 62}, 020303(R) (2001).

\item \label{bib:holland} P.\ R.\ Holland, {\it The Quantum Theory of Motion}
(Cambridge University Press, Cambridge, 1993) pp.\ 86--87, 141--146.

\item \label{bib:prd25} E.\ R.\ Floyd,  Phys.\ Rev.\ {\bf D 25}, 1547 (1982).

\item \label{bib:poirier} B.\ Poirier, J.\ Chem.\ Phys.\ {\bf 121}, 4501 (2004).

\item \label{bib:wyatt} R.\ E.\ Wyatt, {\it Quantum Dynamics with Trajectories:
Introduction to Quantum Hydrodynamics} (Springer, New York, 2005).

\item \label{bib:ch} J.\ F.\ Clauser and M.\ A.\ Horne, Phys.\ Rev.\ {\bfseries D
10}, 526 ((1974).

\item \label{bib:ijmpa15} E.\ R.\ Floyd, Int.\ J.\ Mod.\ Phys.\ {\bfseries A 15},
1363 (2000), quant-ph/9907092.

\item \label{bib:peres} A.\ Peres, Amer.\ J.\ Phys.\ {\bfseries 48}, 552 (1980).

\item \label{bib:bohm} D.\ Bohm, Phys.\ Rev.\ {\bfseries 85}, 166 (1953).

\item \label{bib:ghose} P.\ Ghose, Adv.\ Sc.\ Lett., vol 2, pp 97-99 (2009), arXiv:0905.2037v1.

\item \label{bib:cramer} J.\ G.\ Cramer. Rev.\ Mod.\ Phys.\ {\bfseries 58}, 647-688,
July (1986).

\item \label{bib:bh} D.\ Bohm and B.\ J.\ Hiley, {\itshape The Undivided Universe}
(London, Routledge, 1993).

\end{enumerate}

\bigskip

\clearpage

\noindent {\bf Figure Captions}

\bigskip

\noindent Fig.\ 1. Motion, $x(t)$, of the epr-molecule for $\tau=0$, $A=1$, $B=0.5$,
$k=\pi/2$ and $\beta = 0$ as a solid line and for $\beta = \pi$ as a dashed line.

\bigskip

\noindent Fig.\ 2. Motion of the epr-molecule, $x(t)$, for $\tau=0$, $A=1$, $B=0.5$,
$k=\pi/2$ and $\beta = 0,\pi/4,\pi/2, \cdots,7\pi/4$ for a set of trajectories. All
trajectories are displayed as solid lines.

\end{document}